# INDUCTIVE-ADDER KICKER MODULATOR FOR DARHT-2

E. G. Cook, B.S. Lee, S.A. Hawkins, F.V. Allen, B.C. Hickman, H.C. Kirbie
Lawrence Livermore National Laboratory
C.A. Brooksby – Bechtel Nevada

*Abstract*

An all solid-state kicker modulator for the Dual-Axis Radiographic Hydrodynamic Test facility (DARHT-2) has been designed and tested. This kicker modulator uses multiple solid-state modules stacked in an inductive-adder configuration where the energy is switched into each section of the adder by a parallel array of MOSFETs. The modulator features very fast rise and fall times, pulse width agility and a high pulse-repetition rate in burst mode. The modulator can drive a 50 cable with voltages up to 20 kV and can be easily configured for either positive or negative polarity. The presentation will include test data collected from both the ETA II accelerator kicker and resistive dummy loads.

## 1 BACKGROUND

The DARHT-2 accelerator facility is designed to generate 1 kA electron beam pulses of 2μs duration. The LLNL designed fast kicker, based on cylindrical electromagnetic stripline structures, cleaves four short pulses out of this long pulse. The requirements for the modulator that drives this kicker are listed in Table 1.

Table 1. Performance Requirements

| Parameter | Requirement |
|---|---|
| Output Voltage | ±20kV into 50 |
| Voltage Rise/Falltime | 10ns (10-90%) |
| Flattop Pulsewidth | 16ns–200ns (continuously adjustable) |
| Burst Rate | 4 pulses @1.6MHz(~600ns between leading edges) |

A ±10kV modulator design based on planar triodes was originally used for this application [1]. While the hard-tube performance was very good, concerns regarding future availability and reliability of these devices led to consideration of a solid-state replacement. Personnel within this program had developed considerable expertise with parallel and series arrays of power MOSFETs during the successful design and testing of the Advanced Radiograph Machine (ARM) modulator, a high power pulser developed to show feasibility of solid-state modulators for driving induction accelerators. While ARM was designed for higher voltages and currents than required by the kicker, its requirements for rise and fall times were also significantly slower. After consideration of various circuit topologies, the adder configuration used by ARM was selected as the baseline for the kicker modulator; MOSFETs were selected as the switching device.

The key parameter in the performance requirement is the minimum pulsewidth of 16ns. As a class of devices, 1kV rated MOSFETs have demonstrated the required rise and falltime; however, the critical information needed was to determine whether MOSFETs are capable of switching significant current while simultaneously achieving the required minimum pulsewidth. Device datasheets do not necessarily provide all the information required to make a definitive decision: testing is essential.

## 2 DEVICE EVALUATION AND SELECTION

In order to use a reasonable number of devices, only MOSFETs capable of operation at voltages of 800 volts were evaluated. The evaluation circuit is a series circuit consisting of a low inductance DC capacitor bank, a resistive load, and the MOSFET. Devices were evaluated on the basis of switching speed at various peak currents, waveshapes, minimum output pulsewidth, and ease of triggering. Extensive testing of many devices from several vendors produced several that were acceptable and led to the selection of the APT1001RBVR. During testing, this device exhibited the cleanest rise and fall waveshapes and met the pulsewidth, risetime, and falltime requirements. We were also able to measure a peak current of ~35 amperes before seeing an unacceptable drain-source voltage drop (we arbitrarily chose a voltage drop of < 20 volts during conduction of the current pulse as our acceptance criteria). The APT1001RBVR has a 1000V maximum drain to source rating, an average current rating of 10A, and a pulsed current rating of 40A.

During the early testing of MOSFETs, it became apparent that the MOSFET gate drive circuit was also an essential element in achieving the best performance from the individual devices. The coupling between the drive circuit and the MOSFET had to have very low loop inductance as the peak drive current required to achieve fast switching performance was on the order of tens of amperes. Even the devices within the gate drive circuit had to be very fast and have short turn-on and turn-off delay times. An early decision was that each MOSFET would require its own dedicated gate drive. A simplified schematic of the drive circuit is shown in Fig. 1. The input device of the gate drive has a level-shifting TTL input circuit internally coupled to a MOSFET totem pole output. This circuit drives a fast, high current MOSFET (peak current ±20 amperes) totem pole device which drives the gate of the power MOSFET (capacitive load) to turn it on and sinks current from the MOSFET to turn it off. The gate drive circuit components require a dc voltage of ~ 15 volts.

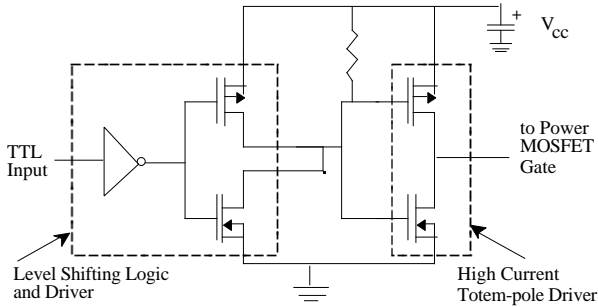

Fig. 1 Simplified Schematic of MOSFET Drive Circuit

## 3 CIRCUIT TOPOLOGY

In the adder configuration shown in Fig. 2, the secondary windings of a number of 1:1 pulse transformers are connected in series. Typically for fast pulse applications, both the primary and secondary winding consists of a single turn to keep the leakage inductance small. In this configuration, the output voltage on the secondary winding is the sum of all the voltages appearing on the primary windings. The source impedance of the MOSFET array and the DC capacitor bank must be very low (<<1 ) to be able to provide the total secondary current, any additional current loads in the primary circuit, plus the magnetization current for the transformer core.

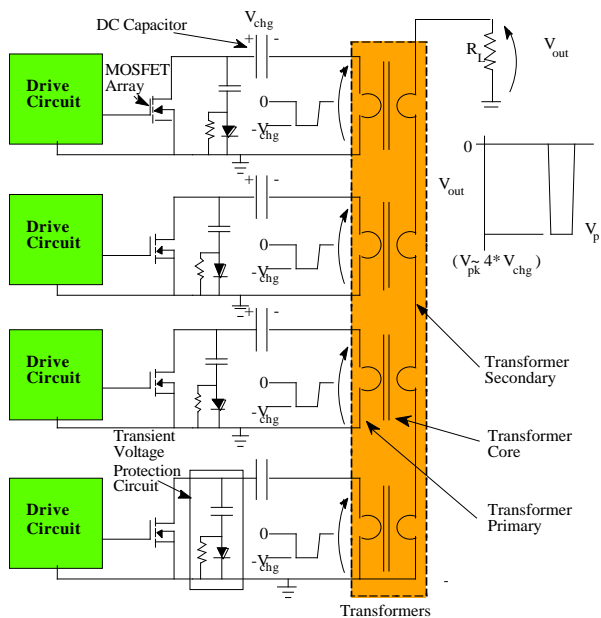

Fig. 2 Simplified Schematic of Adder Circuit

The layout for this circuit is important as it is necessary to mimimize total loop inductance – it doesn't take much inductance to affect performance when the switched di/dt is greater than 40kA/μs.

The MOSFETs shown in Fig. 2 have their source lead connected to ground. This is chosen so that all the gate drive circuits are also ground referenced, thereby eliminating the need for floating and isolated power supplies. The pulse power ground and the drive circuit ground have a common point at the MOSFET source but otherwise do not share common current paths thereby reducing switching transients being coupled into the low level gate drive circuits.

Overvoltage transients can be generated by energy stored in the stray loop inductance, energy stored in the transformer primary, and/or voltage coupled into the primary circuit from the secondary (usually due to trigger timing differences in stages of the adder). Transient protection for the MOSFETs is provided by the series combination of snubber capacitor and diode tightly coupled to the MOSFET. The capacitor is initially charged to the same voltage as the DC capacitor bank. When the MOSFET is turning on, the diode prevents the snubber capacitor from discharging through the MOSFET. As the MOSFET turns off, transient voltages that may exceed the voltage on the snubber capacitor turns the diode on so that the capacitor can absorb the energy. The parallel resistor allows the excess capacitor voltage to discharge into the DC capacitor between bursts. Good performance of the overvoltage circuit requires a low inductance capacitor and a diode with a low forward recovery voltage.

Not shown in the simplified circuit layout is the reset circuit for the magnetic cores. The cores require reset so that they do not saturate during a voltage pulse. As this circuit operates in a well defined pulse format, it is not necessary to actively reset the core between pulses. Consequently, a DC reset circuit is used and is implemented by connecting a DC power supply through a large isolation inductor to the ungrounded end of the secondary winding of the adder stack. In the interval between bursts, the reset current will reset and bias the magnetic cores. This approach is simple to incorporate and requires few additional components but has the disadvantage of requiring more magnetic core material in the transformers.

## 4 COMPONENT LAYOUT

The overall circuit is chosen to have 24 MOSFETs per primary circuit (12 per board). This gives a comfortable margin in peak current capability that allows for extra loading in the primary circuit, a reasonable magnetization current, and total load current. The adder transformer is designed to look very much like an accelerator cell of a linear induction accelerator with the primary winding totally enclosing the magnetic core (an annealed and Namlite insulated Metlgas® 2605 S1A tapewound toroid purchased from National/Arnold). A photograph of a MOSFET carrier board connected to a transformer assembly is shown in Fig. 4. The gate drive circuit boards receive their trigger pulses from a single trigger circuit which is connected to the pulse generator by either optical fiber or coaxial cables. A complete adder assembly is stack of transformer assemblies bolted together as shown in Fig. 5. The secondary winding is usually a metal rod that is positioned on the axial centerline of the adder stack. The rod may be grounded at either end of the adder stack to generate an output voltage of either polarity.

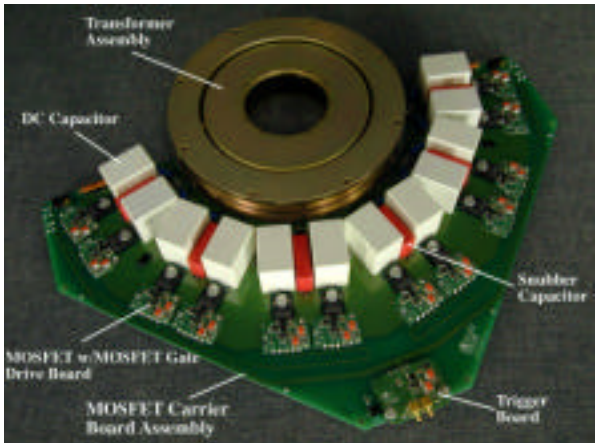

Fig. 4 Transformer Assy. with a MOSFET Carrier Bd.

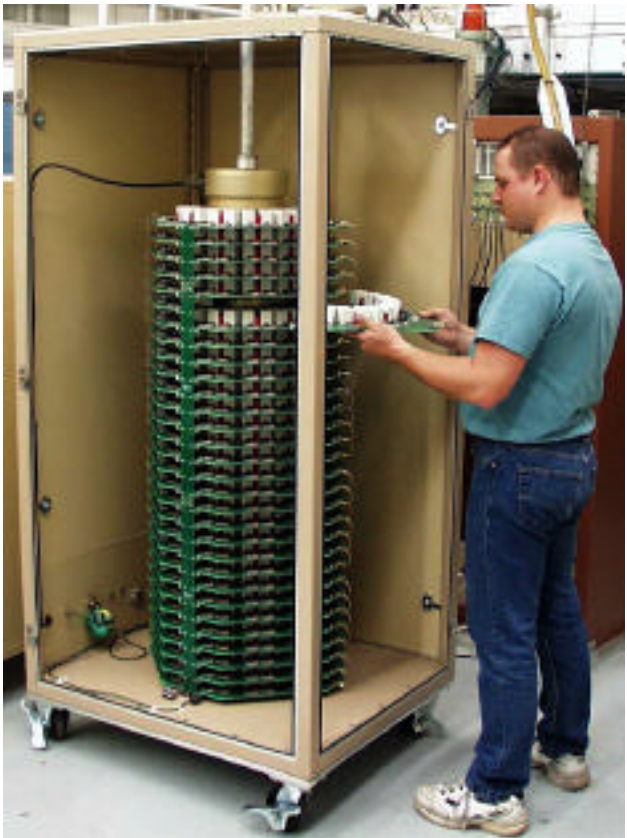

Fig. 5 Complete Kicker Modulator Assembly

## 5 TEST RESULTS

The modulator is undergoing extensive testing into both resistive loads and into the kicker structure used on the ETA II accelerator at LLNL. The modulator has been operated at variable pulsewidths and at burst frequencies exceeding 15 MHz. Fig. 6 is an oscillograph depicting operation on ETA II (with ~500A electron beam current) at ~18kV into 50 Ω (Ch1 is the drain voltage on a single MOSFET and Ch4 is the output current at 100A/div). The four pulse burst in Fig. 7 demonstrates the pulsewidth agility of the modulator at variable burst frequency at an output voltage of ~10kV also into 50 Ω.

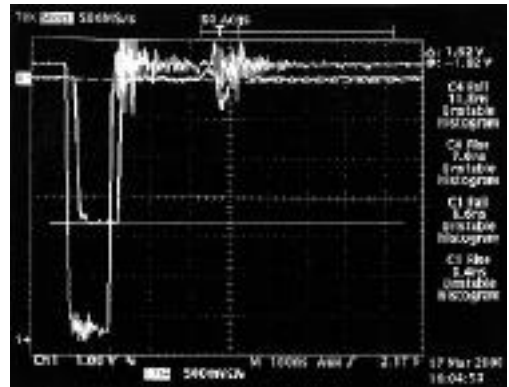

Fig. 6 Operation of Kicker Pulser on ETA II

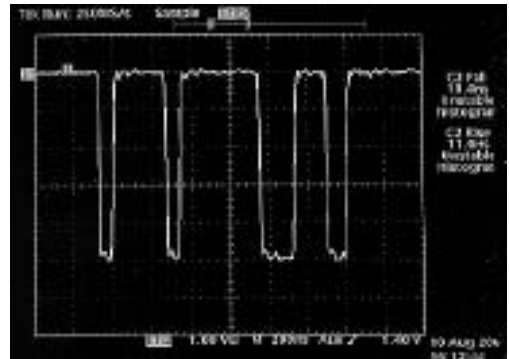

Fig. 7 Four Pulse Burst at 10 kV into 50 Ω Load

## 6 CONCLUSIONS

A fast kicker modulator based on MOSFET switched adder technology has been designed and tested. MOSFET arrays in an adder configuration have demonstrated the ability to generate short duration and very fast risetime and falltime high-voltage pulses.